\documentstyle[12pt]{article}
\newcommand\be{\begin{equation}}
\newcommand\ee{\end{equation}}
\newcommand\bea{\begin{eqnarray}}
\newcommand\eea{\end{eqnarray}}

\newcommand{\fatalpha}{{\bf \alpha \kern -0.44em \alpha}}
\newcommand{\fatsigma}{{\bf \sigma \kern -0.54em \sigma}}
\newcommand{\tpchi}{{\bf \chi \kern -0.35em \chi}}
\newcommand{\llambda}{{\bf \lambda \kern -0.45em \lambda}}

\bibliography{plain}
\title{\bf Quantum discord evolution of three-qubit states under noisy channels }\vspace{20mm}

\author{ M. Mahdian$^a$\thanks{Mahdian@tabrizu.ac.ir},
          R. Yousefjani$^b$\thanks{R.yousefjany@uok.ac.ir}
          and S. Salimi$^b$\thanks{shsalimi@uok.ac.ir}\\
          {$^a$\footnotesize \emph{Department of Theoretical Physics and Astrophysics, University of Tabriz,}}\\
           {\footnotesize \emph{P.O.Box 51664 , Tabriz , Iran}.}\\
            {$^b$\footnotesize \emph{Department of Physics, University of Kurdistan,}}\\
          {\footnotesize \emph{P.O.Box 66177-15175 , Sanandaj,
           Iran}.}
        }

\pagebreak

\vspace{20mm}
\begin{document}
\maketitle \vspace{10mm}
\begin{abstract}
We investigated the dissipative dynamics of quantum discord for correlated qubits under Markovian environments. The basic idea in the present scheme is that quantum discord is more general, and possibly more robust and fundamental, than entanglement. We provide three initially correlated qubits in pure Greenberger-Horne-Zeilinger (GHZ) or W state and analyse the time evolution of the quantum discord under various dissipative channels such as: Pauli channels $\sigma_{x}$, $\sigma_{y}$, and $\sigma_{z}$, as well as depolarising channels.
Surprisingly, we find that under the action of Pauli channel $\sigma_{x}$, the quantum discord of GHZ state is not affected by decoherence.
For the remaining dissipative channels, the W state is more robust than the GHZ state against decoherence.
Moreover, we compare the dynamics of entanglement with that of the quantum discord under the conditions in which disentanglement occurs and show that quantum discord is more robust than entanglement except for phase flip coupling of the three qubits system to the environment.
\end{abstract}
\section{Introduction}
One of the most remarkable properties of quantum mechanics is represented by the quantum correlation.
Entanglement, which is a prominent feature of quantum correlation plays an important role in quantum computing and informational processing \cite{52,53,54,55}. Recently, it has been perceived that entanglement is not the only kind of quantum correlation. In this context, others suitable measures of quantum correlations such as quantum discord \cite{20}, quantum deficit \cite{60,61,62}, quantumness of correlations \cite{63} and quantum dissonance \cite{64} have been proposed. Among them, quantum discord as a measure that is based on the difference between two quantum extensions of classically equivalent concepts has received considerable attention. This measure which quantifies the all nonclassical correlations between parts of a quantum system, actually supplements the measure of entanglement. For pure entangled states quantum discord coincides with the entropy of entanglement. It can also be nonzero for
some mixed separable state.
It is worth mentioning that quantum discord is related to other concepts such as Maxwell's demons \cite{31,32}, quantum phase transitions \cite{33,34}, completely positive maps \cite{35}, and relative entropy \cite{36}.
Moreover, the characteristics of quantum discord have been studied in some physical models and information processing. It was shown that quantum discord can be considered as a more universal quantum resource than quantum entanglement in some sense. It offers new prospects
for quantum information processing \cite{22,24,25,26,27}.
Studying of the quantum discord evolution exposed to noisy environments has led to the surprising result that it may obviously differ from
entanglement evolution. In fact, the unavoidable interaction of the systems with their environment, based on completely positive quantum dynamical semi groups, can be modelled by means of the noisy quantum channels. Quantum channels are completely positive and trace preserving maps between spaces of operators. The study of those can be broadly divided into Markovian and non-Markovian channels depend on the interaction of the system and environment. Markovian channels describes memoryless environments. The prototype of it is given by a quantum dynamical semi group, that is by solving a master equation for the reduced density matrix with Lindblad structure \cite{15,37,38}. For non-Markovian channels, environmental memory plays an important role, so the master equations describing their dynamics are often complicated integro-differential equations which are rarely exactly solvable \cite{37,38}.\\ \\ When a system of qubits with quantum correlation is exposed to noisy channels disentanglement can occur suddenly, but the quantum discord mostly decays in the asymptotic time \cite{57,58,59, 65,66}. This points to a controversial fact that quantum discord may be more robust against decoherence than entanglement. Hence quantum algorithms that are based only on the quantum discord correlations can be more robust than those based on entanglement. The aim of this paper is to illustrate the mentioned subject for a system of three-qubit which is initially prepared in pure state by Greenberger-Horne-Zeilinger (GHZ) \cite{44} or W \cite{45} state as
\begin{eqnarray} \label{1-1}
|\psi ^{GHZ}\rangle&=&\frac{1}{\sqrt{2}}(|000\rangle+|111\rangle)\cr
|\psi ^{W}\rangle&=&\frac{1}{2}(|100\rangle+|010\rangle+\sqrt{2}|001\rangle).
\end{eqnarray}
Jung et al. \cite{1} showed that these pure states will be mixed due to transmission through some of the common channels for qubits.
Moreover, it was shown that under these noisy channels the sudden death of entanglement of three qubits occurs in a finite time \cite{51}.
The question is, what happens to quantum discord under the same conditions in which disentanglement can occur?\\ In reply to this question we lead to remarkable results. The decoherence induced by the Pauli channel $\sigma_{x}$, can not affect the quantum discord of GHZ state contrary to the W state.
We also show that the W state lose less quantum discord than the GHZ state due to transmission through the Pauli channels $\sigma_{y}$, $\sigma_{z}$ and the depolarising channel. Comparison of entanglement and quantum discord demonstrates that for the Pauli channels $\sigma_{x}$ and $\sigma_{y}$ and the depolarising channel, quantum discord is more robust against decoherence than entanglement.
\\ \\The organisation of this paper is as follows:
Section $2$ is devoted to the necessary theoretical background to describe the time evolution of the system and introduces the global quantum discord. Evolution of quantum discord in transmission through the Pauli and the depolarising channels for the GHZ and W states is calculated in section $3$ and $4$, respectively. Finally, we summarise our results.


\section{Method}
\subsection{Time evolution of states under Markovian channels}
In an open quantum system the prototype of a Markov process is given by a quantum dynamical semigroup of a completely positive and trace preserving map $\Phi(t)=\exp[\pounds t]$, $t\geq 0$. In this case, quantum evolution of the system is given by the solution of a Markovian master equation with Lindblad structure for the reduced density matrix \cite{15},
\begin{eqnarray} \label{1-2}
\frac{d}{dt}\rho(t)&=&\pounds \rho(t)\cr
&=& -i[H_{s},\rho(t)]+\sum_{i}L_{i}\rho(t)L_{i}^{\dag}-\frac{1}{2}\{L_{i}^{\dag}L_{i},\rho(t)\}.
\end{eqnarray}
For any Pauli channel $\sigma_{\alpha}$ ($\alpha=x,y,z$), the decoherence dynamic is described by  Lindblad operators $L_{A_{j},\alpha}\equiv\sqrt{\kappa_{A_{j},\alpha}}\,\sigma_{\alpha}^{A_{j}}$, $j=1,2,3$, which act independently upon the $j$-th qubit.
In these operators, $\sigma_{\alpha}^{A_{j}}$ denote the Pauli matrices of the $j$-th qubit and the constants $\kappa_{A_{j},\alpha}$ are relaxation rates. For depolarising channel nine of these operators are needed.
Here we assume that the Hamiltonian of the system is zero $H_{s}=0$, and the strength of the coupling between each of the qubits and channels is equal. Recently, Jung et al. \cite{1} analysed time evolution of three qubit GHZ and W states in the presence of noisy channels. Here we use their results.


\subsection{Quantum discord}
For a bipartite system AB quantum discord is given by  \cite{20}
\begin{eqnarray} \label{2-2}
\textit{D}(\rho^{AB})=\textit{I}(\rho^{AB})- \textit{C}(\rho^{AB}),
\end{eqnarray}
where $\textit{I}(\rho^{AB})=S(\rho^{A})+S(\rho^{B})-S(\rho^{AB})$,
is quantum mutual information which includes the total correlation between A and B. The last section on the right represents classical correlation $\textit{C}(\rho^{AB})=max_{\{\Pi_{k}\}}[S(\rho^{A})-S(\rho^{AB}|\{\Pi_{k}\})]$ with $S(\rho)=-Tr[\rho \log_{2}\rho]$ as the von-Neumann entropy. Notice that the maximum is taken over the set of projective measurements $\{\Pi_{k}\}$ \cite{4}.
\\By definition the conditional density operator $\rho^{AB}_{k}=\frac{1}{p_{k}}\{(I^{A}\otimes \Pi_{k}^{B})\rho^{AB}(I^{A}\otimes \Pi_{k}^{B})\}$ with $p_{k}=Tr[(I^{A}\otimes \Pi_{k}^{B})\rho^{AB}]$ as the probability of obtaining the outcome $k$. We can define the conditional entropy of $A$ as
$S(\rho^{AB}|\{\Pi_{k}\})=\sum_{k}p_{k}S(\rho_{k}^{A})$. This entropy includes the knowledge of subsystem $B$, with $\rho_{k}^{A}=Tr_{B}[\rho_{k}^{AB}]$ and $S(\rho_{k}^{A})=S(\rho_{k}^{AB})$.
It has been shown that $\textit{D}(\rho^{AB})\geq 0$ with the equal sign, only for classical correlation \cite{5}.
\\Very recently, Rulli et al. \cite{3} have proposed a global measure of quantum discord based on a systematic extension of the bipartite quantum discord. Global quantum discord (GQD) which satisfy the basic requirements of a correlation function, for an arbitrary multipartite state $\rho^{A_{1}...A_{N}}$ under a set of local measurement $\{\Pi_{j}^{A_{1}}\otimes ... \otimes \Pi_{j}^{A_{N}}\}$ is defined as
\begin{eqnarray} \label{2-3}
\textit{D}(\rho^{A_{1}...A_{N}})=\min_{\{\Pi_{k}\}}\,[S(\rho^{A_{1}...A_{N}}\|\Phi(\rho^{A_{1}...A_{N}}))-\sum_{j=1}^{N}S(\rho^{A_{j}}\|\Phi_{j}(\rho^{A_{j}}))].
\end{eqnarray}
Where $\Phi_{j}(\rho^{A_{j}})=\sum_{i}\Pi_{i}^{A_{j}}\rho^{A_{j}}\Pi_{i}^{A_{j}}$ and $\Phi(\rho^{A_{1}...A_{N}})=\sum_{k}\Pi_{k}\rho^{A_{1}...A_{N}}\Pi_{k}$ with $\Pi_{k}=\Pi_{j_{1}}^{A_{1}}\otimes ... \otimes\Pi_{j_{N}}^{A_{N}}$ and $k$ denoting the index string ($j_{1}...j_{N}$).
We could eliminate dependence on measurement by minimising the set of projectors $\{\Pi_{j_{1}}^{A_{1}}, ... ,\Pi_{j_{N}}^{A_{N}}\}$.
\\With these remarks about the global quantum discord (4), one can describe the time evolution of the quantum discord for three-qubit GHZ and W  states when they are passed through a noisy channel. By selecting a set of von-Neumann measurements as
\begin{eqnarray}
\Pi_{1}^{A_{j}}=\left(\matrix{\cos^{2}(\frac{\theta_{j}}{2})& \,\,\,\,\,\,\,\,\,e^{i\varphi_{j}}\cos(\frac{\theta_{j}}{2})\sin(\frac{\theta_{j}}{2})\cr e^{-i\varphi_{j}}\cos(\frac{\theta_{j}}{2})\sin(\frac{\theta_{j}}{2})& \,\,\,\,\,\,\,\,\,\sin^{2}(\frac{\theta_{j}}{2})}\right), \cr \cr \cr
\Pi_{2}^{A_{j}}=\left(\matrix{\sin^{2}(\frac{\theta_{j}}{2})& -e^{-i\varphi_{j}}\cos(\frac{\theta_{j}}{2})\sin(\frac{\theta_{j}}{2})\cr -e^{i\varphi_{j}}\cos(\frac{\theta_{j}}{2})\sin(\frac{\theta_{j}}{2})&\cos^{2}(\frac{\theta_{j}}{2})}\right),
\end{eqnarray}
the quantum discord for $\rho=\rho^{GHZ}(t)$ or $\rho^{W}(t)$ can be obtained from
\begin{eqnarray} \label{2-3}
\textit{D}(\rho)=\min_{\{\theta_{j},\varphi_{j}\}}\,[S(\rho\|\Phi(\rho))-\sum_{j=1}^{3}S(\rho^{A_{j}}\|\Phi_{j}(\rho^{A_{j}}))].
\end{eqnarray}
Where $\theta_{j}\in[0,\pi)$ and $\varphi_{j}\in[0,2\pi)$, for $j=1,2,3$, are azimuthal and polar angles, respectively.


\section{Quantum discord of three-qubit system with initial GHZ state under Markovian noise channel}
\subsection{Pauli channel $\sigma_{x}$}
When a three-qubit system with initial GHZ state is coupled to a shift-flip noise channel, in which each qubit is coupled to its own
channel, the time evolution is obtained by the solution of the master equation (2) with Lindbald operators  $L_{j,x}\equiv\sqrt{\kappa_{j,x}}\,\sigma_{x}^{A_{j}}, (j=1,2,3$). For this coupling, the master equation reduces to 8 diagonal and 28 off-diagonal coupled linear differential equations. So after transmission of the GHZ through the Pauli channel $\sigma_{x}$ the density matrix is given by \cite{1}
\begin{eqnarray}\label{8-3}
\rho_{x}^{GHZ}(t)=\frac{1}{8}\left(\matrix{\alpha_{+}&0&0&0&0&0&0&\alpha_{+}\cr 0&\alpha_{-}&0&0&0&0&\alpha_{-}&0 \cr 0&0&\alpha_{-}&0&0&\alpha_{-}&0&0 \cr 0&0&0&\alpha_{-}&\alpha_{-}&0&0&0 \cr 0&0&0&\alpha_{-}&\alpha_{-}&0&0&0 \cr 0&0&\alpha_{-}&0&0&\alpha_{-}&0&0 \cr 0&\alpha_{-}&0&0&0&0&\alpha_{-}&0 \cr \alpha_{+}&0&0&0&0&0&0&\alpha_{+}}\right),
\end{eqnarray}
with
\begin{eqnarray}\label{9-3}
\alpha_{+}&\equiv & 1+3e^{-4\kappa t},\cr
\alpha_{-}&\equiv & 1-e^{-4\kappa t}.
\end{eqnarray}
For this case, the lower bound to the concurrence is \cite{51}
\begin{eqnarray}\label{10-3}
\tau_{3}(\rho_{x}^{GHZ}(t))=e^{-4\kappa t}.
\end{eqnarray}
In order to find the analytical expression for the quantum discord with $\rho_{x}^{GHZ}(t)$ we must consider equation (6).
By tracing out two qubits, the one qubit density matrices representing the individual subsystems are proportional to the identity operator
\begin{eqnarray} \label{12-3}
\rho_{x}^{(A_{1})}(t)=\rho_{x}^{(A_{2})}(t)=\rho_{x}^{(A_{3})}(t)=\frac{\alpha_{+}+3\alpha_{-}}{8}I.
\end{eqnarray}
Therefore, we have $S(\rho_{x}^{(j)}(t)\| \Phi_{j}(\rho_{x}^{(j)}(t)))=0$ $(j=1,2,3)$ and the equation(6) reduces to
\begin{eqnarray} \label{13-3}
\textit{D}(\rho_{x}^{GHZ}(t))&=&\min_{\{\theta_{j},\varphi_{j}\}}\{ S(\rho_{x}^{GHZ}(t)\Vert \Phi(\rho_{x}^{GHZ}(t)))\}\cr
&=&\min_{\{\theta_{j},\varphi_{j}\}}\{ S(\Phi(\rho_{x}^{GHZ}(t)))-S(\rho_{x}^{GHZ}(t))\}.
\end{eqnarray}
The entropy $S(\rho_{x}^{GHZ}(t))$ can be obtained as
\begin{eqnarray} \label{13-3}
S(\rho_{x}^{GHZ}(t))=2-\frac{3\alpha_{-}}{4}\log_{2}\alpha_{-}-\frac{\alpha_{+}}{4}\log_{2}\alpha_{+}.
\end{eqnarray}
In order to obtain the maximum classical correlation among the parts of $\rho_{x}^{GHZ}(t)$, by varying the angles $\theta_{j}$ and $\varphi_{j}$, we must find the measurement bases that minimise $\textit{D}(\rho_{x}^{GHZ}(t))$. After some calculation we  have perceived that, by adopting local measurements in the $\sigma_{z}$ eigenbases for each particle, the value of quantum discord will be minimised. So the von-Neumann entropy, after completing such measurements, can be written as
\begin{eqnarray} \label{13-3}
S(\Phi(\rho_{x}^{GHZ}(t)))=3-\frac{3\alpha_{-}}{4}\log_{2}\alpha_{-}-\frac{\alpha_{+}}{4}\log_{2}\alpha_{+}.
\end{eqnarray}
By substituting equations (12) and (13) into equation (11) quantum discord is readily found to be
\begin{eqnarray} \label{14-3}
\emph{D}(\rho_{x}^{GHZ}(t))=1.
\end{eqnarray}
As can be seen, the Pauli channel $\sigma_{x}$ can not change the quantum discord of the three qubit systems with the initial GHZ state.
We have plotted the dynamic evolution of the quantum discord for density matrix (7) versus the dimensionless scaled time $\kappa t$ in Fig. 1 (dashed violet ).


\subsection{Pauli channel $\sigma_{y}$}
If the three-qubit GHZ state are transmitted through the Pauli channel $\sigma_{y}$, its density matrix takes the following form \cite{1}
\begin{eqnarray}\label{16-3}
\hspace{-7mm}
\rho_{y}^{GHZ}(t)=\frac{1}{8}\left(\matrix{\alpha_{+}&0&0&0&0&0&0&\beta_{1}\cr 0&\alpha_{-}&0&0&0&0&-\beta_{2}&0 \cr 0&0&\alpha_{-}&0&0&-\beta_{2}&0&0 \cr 0&0&0&\alpha_{-}&-\beta_{2}&0&0&0 \cr 0&0&0&-\beta_{2}&\alpha_{-}&0&0&0 \cr 0&0&-\beta_{2}&0&0&\alpha_{-}&0&0 \cr 0&-\beta_{2}&0&0&0&0&\alpha_{-}&0 \cr \beta_{1}&0&0&0&0&0&0&\alpha_{+}}\right),
\end{eqnarray}
where $\alpha_{\pm}$ are given in equation (8) and, $\beta_{1}$ and $\beta_{2}$ are defined as
\begin{eqnarray}\label{17-3}
\beta_{1}&\equiv&3e^{-2\kappa t}+e^{-6\kappa t},\cr
\beta_{2}&\equiv&e^{-2\kappa t}-e^{-6\kappa t}.
\end{eqnarray}
For this matrix, the entanglement vanishes after some finite time due to the condition \cite{51}
\begin{eqnarray}\label{18-3}
\tau_{3}(\rho_{y}^{GHZ}(t))=max\{0,\frac{1}{4}(3e^{-2\kappa t}+e^{-4\kappa t}+e^{-6\kappa t}-1)\}.
\end{eqnarray}
The reduced density matrices of subsystem of (15), is the same as equation (10). Hence, the quantum discord becomes
\begin{eqnarray} \label{20-3}
\textit{D}(\rho_{y}^{GHZ}(t))=\min_{\{\theta_{j},\varphi_{j}\}}\{ S(\Phi(\rho_{y}^{GHZ}(t)))-S(\rho_{y}^{GHZ}(t))\}.
\end{eqnarray}
When the projective measurement in the eigenprojectors of $\sigma_{z}$ is performed on any one of the remaining qubits of $\rho_{y}^{GHZ}(t)$, the minimum quantum discord is obtained as
\begin{eqnarray} \label{21-3}
\textit{D}(\rho_{y}^{GHZ}(t))&=&\frac{(\alpha_{+}-\beta_{1})}{8}\log_{2}(\alpha_{+}-\beta_{1})+\frac{(\alpha_{+}+\beta_{1})}{8}\log_{2}(\alpha_{+}+\beta_{1})\cr
&+&\frac{3(\alpha_{-}-\beta_{2})}{8}\log_{2}(\alpha_{-}-\beta_{2})+\frac{3(\alpha_{-}+\beta_{2})}{8}\log_{2}(\alpha_{-}+\beta_{2})\cr
&-&\frac{\alpha_{+}}{4}\log_{2}(\alpha_{+})-\frac{3\alpha_{-}}{4}\log_{2}(\alpha_{-}).
\end{eqnarray}
Quantum discord is reduced due to the Pauli noisy channel $\sigma_{y}$ and disappears with low speed, as shown in Fig. 1 (solid blue ).


\subsection{Pauli channel $\sigma_{z}$}
Transmission of the GHZ state through the Pauli channel $\sigma_{z}$ result in the master equation (2) reduces to 8 diagonal and 28 off-diagonal first order differential equations with a simply trivial solution. So $\rho^{GHZ}(0)=|GHZ \rangle\langle GHZ|$ is evolved into \cite{1}
\begin{eqnarray} \label{1-3}
\hspace{-9mm}\rho_{z}^{GHZ}(t)=\frac{1}{2}(\vert 000 \rangle\langle 000 \vert + \vert 111 \rangle\langle 111 \vert ) +
\frac{1}{2}e^{-6\kappa t}(\vert 000 \rangle\langle 111 \vert + \vert 000 \rangle\langle 111 \vert ). \nonumber \\
\end{eqnarray}
For this mixed state, the lower bound to concurrence is a mono-exponential function of time
\begin{eqnarray} \label{1-3}
\tau_{3}(\rho_{z}^{GHZ}(t))=e^{-6\kappa t}.
\end{eqnarray}
Let us now focus on the quantum discord for $\rho_{z}^{GHZ}(t)$ as defined by (6). After tracing out two qubits, the three reduce density matrices are equal, given by $\rho_{z}^{(A_{1})}(t)=\rho_{z}^{(A_{2})}(t)=\rho_{z}^{(A_{3})}(t)=\frac{I}{2}$. Hence, we have $S(\rho_{z}^{(A_{j})}(t)\Vert \Phi_{j}(\rho_{z}^{(A_{j})}(t)))=0$ ($j=1,2,3$) and the equation (6) reduces to
\begin{eqnarray} \label{5-3}
\textit{D}(\rho_{z}^{GHZ}(t))=\min_{\{\theta_{j},\varphi_{j}\}}\{ S(\Phi(\rho_{z}^{GHZ}(t)))-S(\rho_{z}^{GHZ}(t))\}.
\end{eqnarray}
The von-Neumann entropy of $\rho_{z}^{GHZ}(t)$ is
\begin{eqnarray} \label{5-3}
\hspace{-10mm}S(\rho_{z}^{GHZ}(t))=\{1-\frac{1+e^{-6\kappa t}}{2}\log_{2}(1+e^{-6\kappa t})-\frac{1-e^{-6\kappa t}}{2}\log_{2}(1-e^{-6\kappa t})\}.\nonumber \\
\end{eqnarray}
We have found that for the density matrix (20), measurements in the eigenprojectors of $\sigma_{z}^{j}$, that is boundary values  $\varphi_{j}=0$ and $\theta_{j}=0$, minimise $\Phi(\rho_{z}^{GHZ}(t))$ as
\begin{eqnarray} \label{5-3}
\Phi(\rho_{z}^{GHZ}(t))=\frac{1}{2}\{|+++\rangle\langle +++|+ |---\rangle\langle ---|\},
\end{eqnarray}
which cause $S(\Phi(\rho_{z}^{GHZ}(t)))=1$. In these circumstances, the quantum discord is expressed as
\begin{eqnarray} \label{6-3}
\textit{D}(\rho_{z}^{GHZ}(t))=\frac{1+e^{-6\kappa t}}{2}\log_{2}(1+e^{-6\kappa t})+\frac{1-e^{-6\kappa t}}{2}\log_{2}(1-e^{-6\kappa t}).
\end{eqnarray}
In Fig. 1, we have plotted the dynamic evolution of the quantum discord for the density matrix (20) as a function of $\kappa t$ (solid red). It can be seen that the quantum discord decreases from its maximal value $\textit{D}(\rho_{z}^{GHZ}(t))=1$ and vanishes after some finite time.


\subsection{Depolarising channel }
For depolarising noise which is described by nine Lindblad operators, $L_{j,z}$, $L_{j,x}$ and $L_{j,y}$ ($j=1,2,3$) the state of three-qubit system that were initially described by the GHZ state replaces
\begin{eqnarray}\label{23-3}
\rho_{d}^{GHZ}(t)=\frac{1}{8}\left(\matrix{\tilde{\alpha}_{+}&0&0&0&0&0&0&\gamma\cr 0&\tilde{\alpha}_{-}&0&0&0&0&0&0 \cr 0&0&\tilde{\alpha}_{-}&0&0&0&0&0 \cr 0&0&0&\tilde{\alpha}_{-}&0&0&0&0 \cr 0&0&0&0&\tilde{\alpha}_{-}&0&0&0 \cr 0&0&0&0&0&\tilde{\alpha}_{-}&0&0 \cr 0&0&0&0&0&0&\tilde{\alpha}_{-}&0 \cr \gamma&0&0&0&0&0&0&\tilde{\alpha}_{+}}\right),
\end{eqnarray}
where
\begin{eqnarray}\label{24-3}
\tilde{\alpha}_{+}&\equiv & 1+3e^{-8\kappa t}\cr
\tilde{\alpha}_{-}&\equiv & 1-e^{-8\kappa t}\cr
\gamma&\equiv &4e^{-12\kappa t}.
\end{eqnarray}
Since $\rho_{d}^{(A_{j})}(t)=\frac{\tilde{\alpha}_{+}+3\tilde{\alpha}_{-}}{8}I$, then $S(\rho_{d}^{(A_{j})}(t)\Vert \Phi_{j}(\rho_{d}^{(A_{j})}(t)))=0$. The quantum discord, due to the condition of depolarizing channel may be written as
\begin{eqnarray} \label{27-3}
\textit{D}(\rho_{d}^{GHZ}(t))=\min_{\theta_{j},\varphi_{j}}\{ S(\Phi(\rho_{d}^{GHZ}(t)))-S(\rho_{d}^{GHZ}(t))\}.
\end{eqnarray}
We find that measurements in the eigenprojectors of $\sigma_{z}^{j}$ minimise the quantum discord. After some algebraic calculation it is found that
\begin{eqnarray} \label{28-3}
\textit{D}(\rho_{d}^{GHZ}(t))&=&\frac{(\alpha_{+}+\gamma)}{8}\log_{2}(\alpha_{+}+\gamma)\cr &+&\frac{(\alpha_{+}-\gamma)}{8}\log_{2}(\alpha_{+}-\gamma)-\frac{\alpha_{+}}{4}
\log_{2}(\alpha_{+}).
\end{eqnarray}
The time evolution of quantum discord (6) as a function of $\kappa t$ in the case of the depolarising channel is plotted in Fig. 1 (solid black). Due to the noisy channel, quantum discord decays from its maximum value $\textit{D}(\rho_{z}^{GHZ}(0))=1$ and disappears after some finite time.
\\ \\ As we have pointed out previously, transmission of the GHZ state through the Pauli channel $\sigma_{x}$ contrary to other channels, could not disturb the quantum discord. We also observed that the quantum discord of the initial state due to transmission of the GHZ state through the Pauli channel $\sigma_{y}$ decayed less quickly than $\sigma_{z}$ and the depolarising channel, as shown in Fig. 1.
In order to comparse quantum discord and entanglement we have used the results of reference \cite{51}.
In Fig. 2, the time dependent entanglement of the GHZ state under the noisy channels is shown.
In all cases, one can see decrease of the entanglement due to the noise of the channels.
Clearly, under the dissipative dynamics considered here, except for phase flip, coupling of the three qubits system to the environment, quantum discord is more robust than entanglement.
\\In the next section we will discuss the effects of noisy channels when we prepare three qubits with initial W state.


\section{Quantum discord of three-qubit systems with initial W state under Markovian noise channel}
\subsection{Pauli channels $\sigma_{x}$ and $\sigma_{y}$}
If the three-qubit system is initially prepared in the W state, a similar analysis as in the previous section to compute the quantum discord can be made.
Under the bit flip or bit-phase flip coupling to environment the density matrix of three-qubit W state at time t has the following analytical expression \cite{1}
\begin{eqnarray}\label{8-3}
\hspace{-25mm}
\rho_{\pm}^{W}(t)=\frac{1}{16}\left(\matrix{
2\alpha_{2}&0&0&\pm\sqrt{2}\alpha_{2}&0&\pm\sqrt{2}\alpha_{2}&\pm\alpha_{2}&0 \cr
0&2\alpha_{1}&\sqrt{2}\alpha_{1}&0&\sqrt{2}\alpha_{1}&0&0&\pm\alpha_{3} \cr
0&\sqrt{2}\alpha_{1}&2\beta_{+}&0&\alpha_{1}&0&0&\pm\sqrt{2}\alpha_{3} \cr
\pm\sqrt{2}\alpha_{2}&0&0&2\beta_{-}&0&\alpha_{4}&\sqrt{2}\alpha_{4} &0 \cr
0&\sqrt{2}\alpha_{1}&\alpha_{1}&0&2\beta_{+}&0&0&\pm\sqrt{2}\alpha_{3} \cr
\pm\sqrt{2}\alpha_{2}&0&0&\alpha_{4}&0&2\beta_{-}&\sqrt{2}\alpha_{4}&0 \cr
\pm\alpha_{2}&0&0&\sqrt{2}\alpha_{4}&0&\sqrt{2}\alpha_{4}&2\alpha_{4}&0 \cr
0&\pm\alpha_{3}&\pm\sqrt{2}\alpha_{3}&0&\pm\sqrt{2}\alpha_{3}&0&0&2\alpha_{3}}
\right),\cr
\end{eqnarray}
where the $+$ sign refers to the $\sigma_{x}$ and $-$ to the $\sigma_{y}$ channel, respectively.
The density matrix elements are given by
\begin{eqnarray}\label{8-3}
\alpha_{1}&=&1+e^{-2\kappa t}+ e^{-4\kappa t}+e^{-6\kappa t}\cr
\alpha_{2}&=&1+e^{-2\kappa t}- e^{-4\kappa t}-e^{-6\kappa t}\cr
\alpha_{3}&=&1-e^{-2\kappa t}- e^{-4\kappa t}+e^{-6\kappa t}\cr
\alpha_{4}&=&1-e^{-2\kappa t}+ e^{-4\kappa t}-e^{-6\kappa t}\cr
\beta_{\pm}&=&1\pm e^{-6\kappa t}.
\end{eqnarray}
It worth mentioning that since two density matrices $\rho_{\pm}^{W}(t)$ have the same structure of matrix elements, we expect that the quantum discord dynamic of these density matrices coincide.
After tracing out any two qubits, the reduced density matrices are given by
\begin{eqnarray} \label{12-3}
\rho_{\pm}^{(A_{j})}(t)=\frac{1}{4}\left(\matrix{2+e^{-2\kappa t}&0\cr 0&2-e^{-2\kappa t}}\right)   \,\,\,\,\, j=1,2,3.
\end{eqnarray}
To find the measurement bases that minimise quantum discord, we used the same procedure as in the previous section.
We perceived that the maximum classical correlation for any reduced density matrix (32) was obtained by completing the projective measurements in the eigenprojectors of $\sigma_{z}$.
It leads to $S(\rho_{\pm}^{(A_{j})}(t)\Vert \Phi_{j}(\rho_{\pm}^{(A_{j})}(t)))=0$. The minimum of $\textit{D}(\rho_{x}^{W}(t))$ and $\textit{D}(\rho_{y}^{W}(t))$ is obtained after the two project measurements in the eigneprojectors of $\sigma_{x}$ and $\sigma_{z}$ on $\rho_{x}^{W}(t)$ and $\rho_{y}^{W}(t)$, respectively.
We do not show the achieved analytic expressions for the quantum discord here because its form are not compact. We have plotted the quantum discord for density matrix (30) as a function of the dimensionless scaled time $\kappa t$ at Fig. 3 (dashed violet). Here, we can see that quantum discord due to the transmission through $\sigma_{x}$ and $\sigma_{y}$ channels, as expected, coincide and decrease at a low rate of speed from its initial value $\textit{D}(\rho_{\pm}^{W}(0))=1.5$ and the asymptotic limit select the exact value $0.813$.


\subsection{Pauli channel $\sigma_{z}$}
After transmission through the Pauli channel $\sigma_{z}$, the time evolution of the W state is described by \cite{1}
\begin{eqnarray}\label{8-3}
\rho_{z}^{W}(t)=\frac{1}{4}\left(\matrix{0&0&0&0&0&0&0&0\cr 0&2&\sqrt{2}e^{-4\kappa t}&0&\sqrt{2}e^{-4\kappa t}&0&0&0 \cr 0&\sqrt{2}e^{-4\kappa t}&1&0&e^{-4\kappa t}&0&0&0 \cr 0&0&0&0&0&0&0&0 \cr 0&\sqrt{2}e^{-4\kappa t}&e^{-4\kappa t}&0&1&0&0&0 \cr 0&0&0&0&0&0&0&0 \cr 0&0&0&0&0&0&0&0 \cr 0&0&0&0&0&0&0&0}\right).
\end{eqnarray}
The density matrices representing each of the one qubit subsystem of (33) are given by $\rho_{z}^{(A_{j})}(t)=\frac{1}{4}\{3|0\rangle\langle0|+|1\rangle\langle1|\}$ ($j=1,2,3$). By using the same procedure as above to find the measurement bases that minimise $\textit{D}(\rho_{z}^{W}(t))$, we find that by performing the projective measurement in the eigenprojectors of $\sigma_{z}^{j}$ quantum discord reaches its lowest point. Thus one can verify that $S(\rho_{z}^{(A_{j})}(t)\Vert \Phi_{j}(\rho_{z}^{(A_{j})}(t)))=0$ and $S(\Phi(\rho_{z}^{W}(t)))=\frac{3}{2}$. The von-Neumann entropy of $\rho_{z}^{W}(t)$ density matrix is found as
\begin{eqnarray} \label{27-3}
S(\rho_{z}^{W}(t))&=&\frac{1}{4}(11+e^{-4\kappa t})-\frac{1}{4}(1-e^{-4\kappa t})\log_{2}(1-e^{-4\kappa t})\cr &-&\frac{1}{8}\{(3+e^{-4\kappa t}-\sqrt{1-2e^{-4\kappa t}+17e^{-8\kappa t}})\cr &&\log_{2}(3+e^{-4\kappa t}-\sqrt{1-2e^{-4\kappa t}+17e^{-8\kappa t}})\cr &+&(3+e^{-4\kappa t}+\sqrt{1-2e^{-4\kappa t}+17e^{-8\kappa t}})\cr && \log_{2}(3+e^{-4\kappa t}+\sqrt{1-2e^{-4\kappa t}+17e^{-8\kappa t}})\}.
\end{eqnarray}
Therefore, the expression for quantum discord $\textit{D}(\rho_{z}^{W}(t))$ is given by
\begin{eqnarray} \label{27-3}
\textit{D}(\rho_{z}^{W}(t))&=&\frac{-1}{4}(5+e^{-4\kappa t})+\frac{1}{4}(1-e^{-4\kappa t})\log_{2}(1-e^{-4\kappa t})\cr &+&\frac{1}{8}\{(3+e^{-4\kappa t}-\sqrt{1-2e^{-4\kappa t}+17e^{-8\kappa t}})\cr &&\log_{2}(3+e^{-4\kappa t}-\sqrt{1-2e^{-4\kappa t}+17e^{-8\kappa t}})\cr &+&(3+e^{-4\kappa t}+\sqrt{1-2e^{-4\kappa t}+17e^{-8\kappa t}})\cr &&\log_{2}(3+e^{-4\kappa t}+\sqrt{1-2e^{-4\kappa t}+17e^{-8\kappa t}})\}.
\end{eqnarray}
With the above channels, quantum discord of the selected system with the initial W state disappear under Pauli channel $\sigma_{z}$ after a finite time. This is displayed in Fig. 3 (solid red).


\subsection{Depolarising channel}
When three-qubite W state is transmitted through the depolarising channel, its time evolution is described by \cite{1}
\begin{eqnarray}\label{8-3}
\hspace{-15mm}
\rho_{d}^{W}(t)=\frac{1}{8}\left(\matrix{
\tilde{\alpha}_{2}&0&0&0&0&0&0&0 \cr
0&\tilde{\alpha}_{1}&\sqrt{2}\tilde{\gamma}_{+}&0&\sqrt{2}\tilde{\gamma}_{+}&0&0&0 \cr
0&\sqrt{2}\tilde{\gamma}_{+}&\tilde{\beta}_{+}&0&\tilde{\gamma}_{+}&0&0&0 \cr
0&0&0&\tilde{\beta}_{-}&0&\tilde{\gamma}_{-}&\sqrt{2}\tilde{\gamma}_{-} &0 \cr
0&\sqrt{2}\tilde{\gamma}_{+}&\tilde{\gamma}_{+}&0&\tilde{\beta}_{+}&0&0&0 \cr
0&0&0&\tilde{\gamma}_{-}&0&\tilde{\beta}_{-}&\sqrt{2}\tilde{\gamma}_{-}&0 \cr
0&0&0&\sqrt{2}\tilde{\gamma}_{-}&0&\sqrt{2}\tilde{\gamma}_{-}&\tilde{\alpha}_{4}&0 \cr
0&0&0&0&0&0&0&\tilde{\alpha}_{3}}
\right),\cr
\end{eqnarray}
with
\begin{eqnarray}\label{8-3}
\tilde{\alpha}_{1}&=&1+e^{-4\kappa t}+ e^{-8\kappa t}+e^{-12\kappa t}\cr
\tilde{\alpha}_{2}&=&1+e^{-4\kappa t}- e^{-8\kappa t}-e^{-12\kappa t}\cr
\tilde{\alpha}_{3}&=&1-e^{-4\kappa t}- e^{-8\kappa t}+e^{-12\kappa t}\cr
\tilde{\alpha}_{4}&=&1-e^{-4\kappa t}+ e^{-8\kappa t}-e^{-12\kappa t}\cr
\tilde{\beta}_{\pm}&=&1\pm e^{-12\kappa t}\cr
\tilde{\gamma}_{\pm}&=&e^{-8\kappa t}\pm e^{-12\kappa t}.
\end{eqnarray}
For density matrix (36), the single qubit density matrix is given by
\begin{eqnarray} \label{12-3}
\rho_{d}^{(A_{j})}(t)=\frac{1}{4}\left(\matrix{2+e^{-4\kappa t}&0\cr 0&2-e^{-4\kappa t}}\right) \,\,\, j=1,2,3.
\end{eqnarray}
Our results show that the minimum quantum discord for the depolarising coupling of the three-qubit W state is attained at boundary values $\varphi_{j}=0$ and $\theta_{j}=0$, that is for a local measurement along the eigenstates of Pauli matrix $\sigma_{z}$. The state of the single qubit is not induced by such measurement, therefore $S(\rho_{d}^{(A_{j})}(t)\Vert \Phi_{j}(\rho_{d}^{(A_{j})}(t)))=0$. The von-Neumann entropy of $\rho_{d}^{W}(t)$ after measurement is given explicitly by
\begin{eqnarray} \label{12-3}
S(\Phi(\rho_{d}^{W}(t)))&=&\frac{1}{4}\{18-\tilde{\alpha}_{1}\log_{2}\tilde{\alpha}_{1}-\tilde{\alpha}_{2}\log_{2}\tilde{\alpha}_{2}\cr &-&\tilde{\alpha}_{3}\log_{2}\tilde{\alpha}_{3}-\tilde{\alpha}_{4}\log_{2}\tilde{\alpha}_{4}-\tilde{\beta}_{+}\log_{2}\tilde{\beta}_{+}
-\tilde{\beta}_{-}\log_{2}\tilde{\beta}_{-}\}.
\end{eqnarray}
In this case, the time evolution of quantum discord is obtained as
\begin{eqnarray} \label{12-3}
\hspace{-15mm}
\textit{D}(\rho_{d}^{W}(t))&=&\frac{1}{2}(3-e^{-8\kappa t})-\frac{1}{4}\{\tilde{\beta}_{+}\log_{2}\tilde{\beta}_{+}
+\tilde{\beta}_{-}\log_{2}\tilde{\beta}_{-}\cr&+&\tilde{\alpha}_{1}\log_{2}\tilde{\alpha}_{1}+2\tilde{\alpha}_{2}\log_{2}\tilde{\alpha}_{2} +2\tilde{\alpha}_{3}\log_{2}\tilde{\alpha}_{3}+\tilde{\alpha}_{4}\log_{2}\tilde{\alpha}_{4}\}\cr&+&\frac{1}{8}\{(\tilde{\beta}_{-}-\tilde{\gamma}_{-})\log_{2}(\tilde{\beta}_{-}-\tilde{\gamma}_{-})+
(\tilde{\beta}_{+}-\tilde{\gamma}_{+})\log_{2}(\tilde{\beta}_{+}-\tilde{\gamma}_{+})\}\cr&+&\frac{1}{16}\{(\tilde{\beta}_{+}+\tilde{\gamma}_{+}+\tilde{\alpha}_{1}
-\sqrt{\tilde{\beta}_{+}^{2}+\tilde{\gamma}_{+}(2\tilde{\beta}_{+}+17\tilde{\gamma}_{+})-2\tilde{\alpha}_{1}(\tilde{\beta}_{+}+\tilde{\gamma}_{+})
+\tilde{\alpha}_{1}^{2}})\cr&&\log_{2}(\tilde{\beta}_{+}+\tilde{\gamma}_{+}+\tilde{\alpha}_{1}
-\sqrt{\tilde{\beta}_{+}^{2}+\tilde{\gamma}_{+}(2\tilde{\beta}_{+}+17\tilde{\gamma}_{+})-2\tilde{\alpha}_{1}(\tilde{\beta}_{+}+\tilde{\gamma}_{+})
+\tilde{\alpha}_{1}^{2}})\cr&+&
(\tilde{\beta}_{+}+\tilde{\gamma}_{+}+\tilde{\alpha}_{1}
+\sqrt{\tilde{\beta}_{+}^{2}+\tilde{\gamma}_{+}(2\tilde{\beta}_{+}+17\tilde{\gamma}_{+})-2\tilde{\alpha}_{1}(\tilde{\beta}_{+}+\tilde{\gamma}_{+})
+\tilde{\alpha}_{1}^{2}})\cr&&\log_{2}(\tilde{\beta}_{+}+\tilde{\gamma}_{+}+\tilde{\alpha}_{1}
+\sqrt{\tilde{\beta}_{+}^{2}+\tilde{\gamma}_{+}(2\tilde{\beta}_{+}+17\tilde{\gamma}_{+})-2\tilde{\alpha}_{1}(\tilde{\beta}_{+}+\tilde{\gamma}_{+})
+\tilde{\alpha}_{1}^{2}})\cr&+&
(\tilde{\beta}_{-}+\tilde{\gamma}_{-}+\tilde{\alpha}_{4}
-\sqrt{\tilde{\beta}_{-}^{2}+\tilde{\gamma}_{-}(2\tilde{\beta}_{-}+17\tilde{\gamma}_{-})-2\tilde{\alpha}_{4}(\tilde{\beta}_{-}+\tilde{\gamma}_{-})
+\tilde{\alpha}_{4}^{2}})\cr&&\log_{2}(\tilde{\beta}_{-}+\tilde{\gamma}_{-}+\tilde{\alpha}_{4}
-\sqrt{\tilde{\beta}_{-}^{2}+\tilde{\gamma}_{-}(2\tilde{\beta}_{-}+17\tilde{\gamma}_{-})-2\tilde{\alpha}_{4}(\tilde{\beta}_{-}+\tilde{\gamma}_{-})
+\tilde{\alpha}_{4}^{2}})\cr&+&
(\tilde{\beta}_{-}+\tilde{\gamma}_{-}+\tilde{\alpha}_{4}
+\sqrt{\tilde{\beta}_{-}^{2}+\tilde{\gamma}_{-}(2\tilde{\beta}_{-}+17\tilde{\gamma}_{-})-2\tilde{\alpha}_{4}(\tilde{\beta}_{-}+\tilde{\gamma}_{-})
+\tilde{\alpha}_{4}^{2}})\cr&&\log_{2}(\tilde{\beta}_{-}+\tilde{\gamma}_{-}+\tilde{\alpha}_{4}
+\sqrt{\tilde{\beta}_{-}^{2}+\tilde{\gamma}_{-}(2\tilde{\beta}_{-}+17\tilde{\gamma}_{-})-2\tilde{\alpha}_{4}(\tilde{\beta}_{-}+\tilde{\gamma}_{-})
+\tilde{\alpha}_{4}^{2}})\}. \nonumber \\
\end{eqnarray}
The responses of quantum discord with the initial W state to the different noises as functions of $\kappa t$ are listed at Fig. 3. As in the case of the GHZ state, the quantum discord for the W state decays due to the noisy channels.
When a three-qubit system with initial W or GHZ state is coupled to a depolarising noise channel decay of the quantum discord occurs so fast. This means that the depolarising channel has the most destructive influence on the quantum discord.
However, the quantum discord decreases due to the transmission through the Pauli channels $\sigma_{x}$ or $\sigma_{y}$, but it catches the exact value $0.813$ in the long term as shown in Fig. 3. This figure also exhibits the vanishing of the quantum discord for the initial W state under the phase flip channel.
These results show that GHZ-type quantum discord is more fragile under certain types of environment coupling which can be modelled by means of the Pauli channels $\sigma_{y}$ and $\sigma_{z }$ as well as the depolarising channel as compared to the quantum discord of the W state. However, in the Pauli channel $\sigma_{x}$ the situation becomes completely reversed.
\\In order to show the difference between the entanglement and quantum discord under the various dissipative channels, by using the results of \cite{51} we have plotted  the evolution of entanglement of the W state in Fig. 4. As in the case of the GHZ state, the entanglement for the W state decays exponentially due to the noisy channels. Observe that under the dissipative dynamics considered, except for the phase flip coupling of the three qubits system to the environment, discord is more robust than entanglement.


\section{SUMMARY AND CONCLUSION}
To sum up, we prepared the three-qubit system with the initial state formed by GHZ and W states. We then studied the dynamics of quantum discord under interaction with independent Markovian environments which can be modelled by means of the various noisy channels, namely, the Pauli channels $\sigma_{x}$, $\sigma_{y}$ and $\sigma_{z}$ as well as the depolarising channel. It was clarified that coupling with the bit flip channel could not disorder the quantum discord of three-qubit GHZ state unlike in the W state. In other words, for this noisy channel, GHZ-type quantum discord is always more robust than quantum discord of the W state. In the Pauli channels $\sigma_{y}$, $\sigma_{z}$ and depolarising channel, the W state preserved more quantum discord than the GHZ state in the long term.
\\ Also, we observed that under the dissipative Markovian dynamics considered, except for the Pauli channel $\sigma_{z}$ quantum discord is more robust than entanglement. This points to the fact that quantum discord is another kind of quantum correlation different from entanglement and the absence of entanglement does not necessarily indicate the absence of quantum correlations.


\newpage
\textbf{Figure Caption}
\itemize{}
\item FIG. 1. (Colour online) The quantum discord (6) for the three-qubit system with the initial GHZ state as a function of $\kappa t$, if transmitted through Markovian channels: Pauli channel $\sigma_{x}$ (dashed violet), $\sigma_{y}$ (solid blue), $\sigma_{z}$ (solid red) and depolarising channel (solid black).

\item FIG. 2. (Colour online) The entanglement for the three-qubit system with the initial GHZ state as a function of $\kappa t$, if transmitted through Markovian channels: Pauli channel $\sigma_{x}$ (dashed violet), $\sigma_{y}$ (solid blue), $\sigma_{z}$ (solid red) and depolarising channel (solid black).
\item FIG. 3.  (Colour online) The quantum discord (6) for the three-qubit system with the initial W state as a function of $\kappa t$, if transmitted through Markovian channels: Pauli channel $\sigma_{x}$ and $\sigma_{y}$ (dashed violet), $\sigma_{z}$ (solid red) and depolarising channel (solid black).

\item FIG. 4. (Colour online) The entanglement for the three-qubit system with the initial W state as a function of $\kappa t$, if transmitted through Markovian channels: Pauli channel $\sigma_{x}$ (dashed violet), $\sigma_{y}$ (solid blue), $\sigma_{z}$ (solid red) and depolarising channel (solid black).

\end{document}